\documentclass[colorlinks=true,linkcolor=red,citecolor=blue,urlcolor=blue,
    aps,
    pra,
    onecolumn,
    superscriptaddress,
    showkeys
]{revtex4-2}

\usepackage[utf8]{inputenc}
\usepackage[T1]{fontenc}
\usepackage{amsmath,graphicx,array,float,amsthm,amssymb}
\usepackage{tabularray}
\usepackage{tikz,color,multirow}
\usepackage{indentfirst,orcidlink}
\usepackage{braket}
\usepackage{diagbox}
\usepackage{bbold,soul}
\usepackage[htt]{hyphenat}

\usepackage{hyperref}

\makeatletter
\renewcommand{\p@subsection}{}
\renewcommand{\p@subsubsection}{}
\makeatother

\begin{document}

\title{Machine Learning of Quantum Entanglement from Noisy Measurements}

\author{Artur Czerwinski~\!\!\orcidlink{0000-0003-0625-8339}}
\email{aczerwin@umk.pl}
\affiliation{Institute of Physics, Faculty of Physics, Astronomy and Informatics, Nicolaus Copernicus University in Torun, ul. Grudziadzka 5, 87-100 Torun, Poland}

\author{Maciej Wiśniewski~\!\!\orcidlink{0009-0008-1262-5186}}
%\email{maciekwisniewski3@abs.umk.pl}
\affiliation{Institute of Physics, Faculty of Physics, Astronomy and Informatics, Nicolaus Copernicus University in Torun, ul. Grudziadzka 5, 87-100 Torun, Poland}

\author{Paweł Moszczyński~\!\!\orcidlink{0000-0001-9034-0698}}
%\email{pawel.moszczynski@wat.edu.pl}
\affiliation{Faculty of Cybernetics, Military University of Technology, ul. gen. Sylwestra Kaliskiego 2, 00-908 Warsaw, Poland}

\begin{abstract}
In this work, we investigate the application of Machine Learning (ML) algorithms to the identification and quantitative characterization of quantum entanglement in polarization-entangled photon pairs. The analysis is based on simulated symmetric, informationally complete, positive operator-valued measure (SIC-POVM) measurement data, where each two-qubit state is represented by a 16-dimensional measurement vector corresponding to experimentally accessible coincidence counts. The generated SIC-POVM measurement data include Poissonian shot noise. Several supervised ML algorithms, including Logistic Regression, k-Nearest Neighbors, Decision Trees, Support Vector Machines, and Random Forests, are applied to the classification of separable and entangled states directly from raw measurement data, without explicit density matrix reconstruction or the use of conventional separability criteria. The study additionally explores clustering methods and nonlinear regression techniques for estimating continuous entanglement measures. The obtained results demonstrate that ML methods can achieve very high classification accuracy, even under extremely limited training conditions. These findings indicate that ML may provide an efficient alternative to conventional quantum-state analysis under simulated Poissonian noise conditions.
\end{abstract}


\maketitle

\section{Introduction}

Quantum entanglement is one of the most distinctive and nonclassical features of quantum mechanics. Since its original discussion in the celebrated Einstein--Podolsky--Rosen paradox and the subsequent formulation of Bell inequalities, entanglement has evolved from a conceptual curiosity into a fundamental resource for quantum information science. Today, entanglement plays a central role in numerous quantum technologies, including quantum communication, quantum cryptography, distributed quantum sensing, quantum metrology, and quantum computing \cite{Nielsen2000,Horodecki2009}.

Among various physical platforms capable of generating and manipulating entangled states, photonic systems occupy a particularly important position. Entangled photon pairs can be efficiently generated using spontaneous parametric down-conversion, whose theoretical foundations were formulated in \cite{HongMandel1985} and subsequently developed in the context of polarization two-photon entanglement in type-II down-conversion in \cite{Rubi1994}. The photons prepared in this way can then be transmitted over long distances through optical fibers or free-space optical channels. Their relatively weak interaction with the environment makes them attractive for practical implementations of quantum communication protocols. Polarization-entangled photons have been extensively employed in experimental demonstrations of quantum key distribution, quantum teleportation, entanglement swapping, quantum repeaters, and satellite-based quantum communications \cite{Gisin2007,Pirandola2020}.

The successful implementation of such protocols requires reliable methods for detecting and characterizing entanglement. In experimental settings, this task is commonly performed through quantum state tomography (QST), which enables reconstruction of the density matrix describing the investigated quantum system. Once the density matrix has been obtained, various separability criteria and entanglement measures can be employed to determine whether the state is entangled and to quantify the degree of entanglement \cite{Peres1996,Horodecki1996,hill1997entanglement,Wootters1998}. Although this approach is rigorous and widely used, it may become computationally demanding when large numbers of quantum states must be analyzed or when measurements are affected by statistical noise.

Recent advances in Machine Learning (ML) have created new opportunities for the analysis of quantum systems. In general, ML methods are now used in many areas of the physical sciences and quantum information, including both the analysis of experimental data and the modeling of quantum systems \cite{Carleo2019,Biamonte2017}. A separate research direction concerns the use of ML to support quantum-state tomography with a limited number of measurements or in the presence of noise \cite{ma2024neural}. Particularly close to the present work are applications of ML to entanglement analysis: from entangled-state detection treated as a classification problem \cite{asif2023entanglement,urena2024entanglement}, to the estimation of quantitative entanglement measures without full quantum-state reconstruction \cite{koutny2023deep,huang2022measuring}.

The present work investigates the application of ML methods to the analysis of quantum entanglement in pairs of polarization-entangled photons. The study is based on simulated noisy tomographic data obtained using symmetric informationally complete positive operator-valued measures (SIC-POVMs) \cite{Renes2004,Fuchs2017}. The present analysis accounts for Poissonian photon-counting noise but does not include additional experimental imperfections, such as channel losses, detector efficiencies, or dark counts. The analysis was carried out in two stages. First, two extreme classes of quantum states were considered, namely separable states and maximally entangled Bell states, in order to determine whether the information required to distinguish between these classes is directly encoded in noisy measurement statistics. For this dataset, both supervised classification and unsupervised clustering methods were applied. Subsequently, the analysis was extended to Werner states, which enable the investigation of quantum states exhibiting different degrees of entanglement. For the Werner-state dataset, supervised classification of separable and entangled states was performed, covering the full range from partially entangled to maximally entangled states. In addition, nonlinear regression techniques were employed to estimate continuous entanglement measures directly from measurement statistics.

The central hypothesis of this work is that ML models can learn the relationship between experimentally accessible measurement outcomes and the entanglement properties of quantum states without requiring explicit density matrix reconstruction or the direct application of conventional separability criteria. The considered methodology is therefore based on using noisy SIC-POVM measurement vectors as direct input features (independent variables) for supervised classification and regression models. In this framework, classification is used to distinguish separable and entangled states, whereas regression is used to estimate concurrence as a quantitative measure of entanglement. If successful, such an approach could significantly simplify the analysis of experimental quantum-optical data and provide a computationally efficient alternative to standard tomographic techniques.

The use of ML methods for the identification of entanglement is associated with certain limitations. The performance of such models depends on the representativeness of the training data, the adopted noise model and the range of analyzed quantum states. For example, very good results obtained for clearly separated classes, such as separable and maximally entangled states, do not imply that ML algorithms will achieve the same effectiveness for arbitrary mixed states, especially those located close to the separability boundary. For this reason, the present work analyzes selected ML methods under controlled conditions of simulated SIC-POVM measurements, indicating both their potential and their limitations.

The paper is organized as follows. Section~\ref{formalismsec} briefly introduces the mathematical formalism used throughout the paper, including the density-matrix representation of quantum states. Section~\ref{generationsec} describes the methodology employed to generate the datasets used in the study, while the detailed procedures for data generation are provided in Appendix~\ref{appendixA} and Appendix~\ref{appendixB} for the separable-versus-Bell-state dataset and the Werner-state dataset, respectively. Section~\ref{mlresults} presents the results obtained using the investigated ML algorithms, including classification and regression performance metrics. Finally, the main conclusions of the study are summarized in Section~\ref{conclusions}.

\section{Quantum States and Entanglement Formalism}\label{formalismsec}

\subsection{Density Matrix Representation}

The mathematical description of quantum systems is based on the formalism of Hilbert spaces and linear operators acting upon them. While isolated quantum systems can be described by state vectors, practical quantum-optical experiments often involve incomplete information, statistical mixtures, and interactions with the environment. Consequently, the density matrix formalism provides a more general and experimentally relevant representation of quantum states.

For a pure state represented by a normalized state vector $\ket{\psi}$, the density matrix is defined as \cite{Nielsen2000,blum2012density}
\begin{equation}
\rho = \ket{\psi}\!\bra{\psi}.
\end{equation}

In the case of a statistical mixture of states $\ket{\psi_i}$ occurring with probabilities $p_i$, we define the density matrix as
\begin{equation}
\rho = \sum_i p_i \ket{\psi_i}\!\bra{\psi_i},
\qquad
\sum_i p_i = 1.
\end{equation}

A physically admissible density matrix is Hermitian, positive semidefinite, and normalized to have unit trace \cite{Nielsen2000,blum2012density}
\begin{equation}
\rho^\dagger=\rho,
\qquad
\rho \geq 0,
\qquad\text{and}\qquad
\mathrm{Tr}(\rho)=1.
\end{equation}

In this work, a system of two polarization-encoded qubits is considered, whose Hilbert space has the form
\begin{equation}
\mathcal{H}=\mathcal{H}_A\otimes\mathcal{H}_B.
\end{equation}
The corresponding density matrix is therefore a $4\times4$ matrix.

\subsection{Separable and Entangled States}

A bipartite quantum state is called separable if it can be expressed as

\begin{equation}
\rho_{\mathrm{sep}}
=
\sum_i p_i \,
\rho_i^A \otimes \rho_i^B,
\end{equation}
where $\rho_i^A$ and $\rho_i^B$ are density matrices describing the individual subsystems and $p_i$ form a probability distribution.

States that cannot be represented in this form are called entangled. Such states exhibit nonclassical correlations that cannot be explained by local hidden-variable models and constitute the essential resource for many quantum information protocols.

Determining whether a given density matrix is separable or entangled is generally a nontrivial problem. Several separability criteria have been developed for this purpose. Among them, the Peres--Horodecki criterion based on partial transposition plays a particularly important role. For bipartite systems of dimensions $2\times2$ and $2\times3$, positivity of the partially transposed density matrix provides a necessary and sufficient condition for separability \cite{Peres1996,Horodecki1996}.

Beyond the binary distinction between separable and entangled states, it is often desirable to quantify the amount of entanglement present in a quantum system. Various entanglement measures have been proposed, including concurrence, negativity, and entanglement of formation. In the present work, concurrence is employed as the primary quantitative measure of entanglement.

For two-qubit systems, concurrence is commonly evaluated using the spin-flip construction \cite{hill1997entanglement,Wootters1998}. For an arbitrary two-qubit density matrix $\rho$, the spin-flipped state is defined as

\begin{equation}
\widetilde{\rho}
=
(\sigma_y\otimes\sigma_y)
\rho^*
(\sigma_y\otimes\sigma_y),
\label{eq:spin_flipped_state_main}
\end{equation}
where $\rho^*$ denotes complex conjugation in the computational basis and $\sigma_y$ is the Pauli-$Y$ matrix. The concurrence is then given by
\begin{equation}
C(\rho)
=
\max\left(
0,\lambda_1-\lambda_2-\lambda_3-\lambda_4
\right),
\label{eq:general_concurrence_main}
\end{equation}
where $\lambda_1\geq\lambda_2\geq\lambda_3\geq\lambda_4$ are the square roots of the eigenvalues of $\rho\widetilde{\rho}$. Concurrence assumes values in the interval $0\leq C(\rho)\leq1$, where $C(\rho)=0$ corresponds to a separable state and $C(\rho)=1$ corresponds to a maximally entangled state. A more detailed description of the concurrence calculation is provided in Appendix~\ref{appendixB}.

\subsection{Quantum Measurements and Tomographic Data}\label{sec:SIC-POVMData}

Experimental determination of a density matrix is typically performed using QST frameworks \cite{hradil1997quantum,paris2004quantum}. This procedure reconstructs the density matrix from a sufficiently large set of measurement outcomes \cite{jamiolkowski1983minimal,jamiolkowski2000complete,czerwinski2022selected}.

For a single qubit, four independent measurement operators are required to obtain complete information about the quantum state \cite{james2001measurement}. Consequently, a two-qubit system requires
\begin{equation}
4^2 = 16
\end{equation}
independent measurements.

In this work, the measurement scheme is based on a single-qubit SIC-POVM constructed from the following four normalized state vectors \cite{vrehavcek2004minimal}:
\begin{equation}\label{sickPOVMdef}
\begin{aligned}
|f_1\rangle &=
\begin{pmatrix}
1\\
0
\end{pmatrix},
\\[0.5em]
|f_2\rangle &=
\frac{1}{\sqrt{3}}
\begin{pmatrix}
1\\
0
\end{pmatrix}
+
\sqrt{\frac{2}{3}}
\begin{pmatrix}
0\\
1
\end{pmatrix},
\\[0.5em]
|f_3\rangle &=
\frac{1}{\sqrt{3}}
\begin{pmatrix}
1\\
0
\end{pmatrix}
+
\sqrt{\frac{2}{3}}
e^{i\frac{2\pi}{3}}
\begin{pmatrix}
0\\
1
\end{pmatrix},
\\[0.5em]
|f_4\rangle &=
\frac{1}{\sqrt{3}}
\begin{pmatrix}
1\\
0
\end{pmatrix}
+
\sqrt{\frac{2}{3}}
e^{i\frac{4\pi}{3}}
\begin{pmatrix}
0\\
1
\end{pmatrix},
\end{aligned}
\end{equation}
which are known to form the vertices of a regular tetrahedron in the Bloch sphere.

The corresponding SIC-POVM elements are defined as
\begin{equation}
P_i=\frac{1}{2}|f_i\rangle\langle f_i|,
\qquad i=1,\ldots,4.
\end{equation}

For the two-qubit polarization system, the sixteen measurement operators are constructed as
\begin{equation}
M_{ij}=P_i\otimes P_j,
\qquad i,j\in\{1,2,3,4\}.
\end{equation}

These operators form an informationally complete measurement set and yield a $16$-dimensional vector of experimentally accessible coincidence counts for each investigated quantum state. The central idea investigated in this paper is that such measurement vectors, despite being affected by Poissonian shot noise, may contain sufficient information to determine the entanglement properties of the underlying quantum state. Consequently, ML algorithms may be capable of extracting this information directly from the measurement data, eliminating the need for explicit density-matrix reconstruction. Further implementation details are provided in Appendix~\ref{appendixA}.

\section{Generation of Quantum States and Measurement Data}\label{generationsec}

The performance of ML algorithms strongly depends on the quality, diversity, and physical realism of the training data. Since large experimentally acquired datasets of entangled photonic states are not readily available, the present study relies on numerically generated quantum states and simulated tomographic measurements. The datasets were constructed to incorporate Poissonian photon-counting fluctuations and included quantum states spanning a broad range of geometries and entanglement properties.

For all investigated scenarios, quantum states were represented by two-qubit density matrices corresponding to polarization-entangled photon pairs. Measurement data were generated using SIC-POVMs formalism, yielding experimentally accessible 16-dimensional measurement vectors that include shot noise \cite{hasinoff2021photon,sedziak2020tomography}.

\subsection{Dataset for Classification and Clustering with the Bell Class}\label{IIIA}

The first dataset was designed for the binary classification of quantum states and for unsupervised clustering analysis. Two distinct classes of quantum states were generated: separable states and maximally entangled states.

The ensemble of maximally entangled states was generated by applying randomly selected local unitary transformations to Bell states. Since local unitary operations preserve entanglement, this procedure generated a large collection of geometrically distinct quantum states possessing identical maximal entanglement. Random sampling was subsequently employed to select $10,000$ distinct, representative states from the generated ensemble.

The separable-state dataset was generated independently using tensor products of single-qubit density matrices. Individual qubit states were parametrized using Bloch-sphere coordinates and sampled over a broad range of radii and angular parameters. This procedure produced more than $10^5$ physically valid separable states exhibiting different levels of purity and different orientations within the Bloch sphere. A random subset consisting of $10,000$ distinct states was then selected for further analysis.

For every generated density matrix $\rho$, probabilities associated with sixteen SIC-POVM measurement operators were calculated according to Born's rule. Photon counts were generated using Poisson distributions corresponding to an average count scale of $40,000$ detected photon pairs per measurement setting. Consequently, each quantum state was represented by a noisy $16$-dimensional measurement vector accompanied by a binary label identifying the corresponding class

\begin{equation}
\rho \hspace{0.35cm} \longleftrightarrow \hspace{0.35cm} (n_1,n_2,\ldots,n_{16}, \mathcal{C}^{\mathrm{bin}}),
\end{equation}
where $n_i$ denotes the photon count associated with the $i$-th SIC-POVM outcome and $\mathcal{C}^{\mathrm{bin}}$ takes two possible values: $0$ for separable states and $1$ for entangled states (in this scenario maximally entangled states).

The final classification dataset therefore consisted of $20,000$ measurement vectors, equally divided between separable and entangled states. The detailed procedure for generating quantum states and pseudo-experimental measurement data for classification and clustering is described in Appendix~\ref{appendixA}.

\subsection{Dataset for Classification and Regression with the Werner States}
\label{IIIB}

The second dataset was designed to investigate ML methods for both the classification of separable and entangled states and the quantitative estimation of entanglement from measurement data. Unlike the dataset described in Section~\ref{IIIA}, which consisted exclusively of separable and maximally entangled states, the present dataset spans the entire range of entanglement values.

To generate quantum states with different degrees of entanglement, the family of Werner states~\cite{Werner1989} was adopted as the underlying model. The two-qubit Werner state is defined as
\begin{equation}
\rho_W(p)
=
p\,|\Psi^{-}\rangle\langle\Psi^{-}|
+
(1-p)\frac{\mathbb{I}_4}{4},
\qquad
0\leq p\leq1,
\label{eq:werner_state_main}
\end{equation}
where $|\Psi^{-}\rangle$ is the maximally entangled singlet Bell state, while $\mathbb{I}_4$ denotes the identity operator acting on the two-qubit Hilbert space. These states continuously interpolate between the maximally mixed state and the maximally entangled singlet state through the parameter $p$. They are separable for $p\leq1/3$ and entangled for $p>1/3$, while the limiting case $p=1$ corresponds to the maximally entangled singlet state. In the present study, the parameter was discretized over the interval $[0,1]$ using a uniform step size of $\Delta p=0.02$.

To increase the geometrical diversity of the dataset, independent local unitary transformations were applied to both qubits. Since local unitary operations preserve all entanglement measures, this procedure generated a large ensemble of physically distinct quantum states while maintaining the entanglement associated with the selected value of $p$.

The resulting parameter space contained more than $4.4\times10^6$ valid two-qubit states. From this ensemble, a random subset of $250,000$ density matrices was selected for further analysis. For each state, the concurrence was evaluated and used as the target variable in the regression task. Concurrence was selected because it is one of the most widely used entanglement monotone for two-qubit systems and assumes values in the interval

\begin{equation}
0 \leq C \leq 1,
\end{equation}
where $C=0$ corresponds to separable states and $C=1$ to maximally entangled states.

As in the previous dataset, the probabilities associated with the sixteen SIC-POVM outcomes were calculated for every density matrix according to Born's rule and converted into pseudo-experimental coincidence counts using Poissonian photon-counting statistics.

Each observation in the final dataset was represented by
\begin{equation}
\rho
\hspace{0.35cm}
\longleftrightarrow
\hspace{0.35cm}
(n_1,n_2,\ldots,n_{16},C,\mathcal{C}^{\mathrm{bin}},p),
\end{equation}
where $n_1,\ldots,n_{16}$ denote the simulated SIC-POVM coincidence counts, $p$ is the Werner-state parameter, $C$ is the concurrence, and $\mathcal{C}^{\mathrm{bin}}$ is the binary entanglement label, taking the value $0$ for separable states ($p \leq 1/3$) and $1$ for entangled states ($p > 1/3$). Consequently, the same dataset can be used for both supervised classification and nonlinear regression. The first sixteen components correspond to experimentally accessible measurement outcomes (input variables), whereas the remaining quantities describe the underlying quantum state and provide the target variables for the ML models (output variables).

The resulting dataset enables the investigation of ML models capable of identifying and quantifying quantum entanglement directly from noisy measurement statistics, without reconstructing the density matrix or explicitly evaluating entanglement criteria. A detailed description of the data-generation procedure is provided in Appendix~\ref{appendixB}.

\section{Application of ML algorithms: Results and Analysis}\label{mlresults}

In this section, ML  methods are applied to the generated SIC-POVM measurement data. The analysis is organized according to the two datasets considered in this work. In the first part, classification and clustering of separable and maximally entangled states are investigated, representing clearly separated limiting cases. In the second part, the Werner-state dataset is analyzed, covering states with varying degrees of entanglement. For this dataset, the classification of separable and entangled states and regression-based estimation of concurrence directly from measurement statistics are considered. In each part, the applied methods and the obtained results are presented.

\subsection{Separable and Maximally Entangled States}\label{sec:Separable and Maximally Entangled States}

This section presents the analysis of the dataset containing separable and maximally entangled states. The primary objective was to determine whether noisy SIC-POVM measurement vectors contain sufficient information to distinguish these two well-defined classes of quantum states without explicitly reconstructing the density matrix. To this end, both supervised classification and unsupervised clustering methods were investigated.

\subsubsection{Supervised Classification}

The first ML task was the binary classification of quantum states into separable and maximally entangled classes. The input data consisted of noisy SIC-POVM measurement vectors generated from the dataset described in Section~\ref{IIIA}. For comparison, five classification algorithms were considered: \texttt{Logistic Regression}, \texttt{KNeighborsClassifier}, \texttt{DecisionTreeClassifier}, \texttt{SVC}, and \texttt{RandomForestClassifier} \cite{bishop2006pattern}. The objective was to predict the binary entanglement label, $y=\mathcal{C}^{\mathrm{bin}}$.

The algorithms used in this study were selected to represent different classes of classification approaches. \texttt{Logistic Regression} serves as a linear reference model, allowing us to assess whether the considered classes can be separated using a simple linear decision boundary. \texttt{KNeighborsClassifier} represents a local, instance-based method that classifies states according to the similarity between measurement vectors. \texttt{DecisionTreeClassifier} constructs nonlinear decision boundaries through recursive partitioning of the feature space, whereas \texttt{RandomForestClassifier} provides a more robust ensemble-based extension that reduces the variance and improves the generalization ability of a single decision tree. Finally, \texttt{SVC} was included as a kernel-based classifier capable of modeling highly nonlinear decision boundaries. By comparing these conceptually different approaches, the analysis provides insight not only into their classification performance but also into the geometric structure of the noisy SIC-POVM measurement space. In particular, the relative performance of linear, local, kernel-based, and ensemble methods helps identify the complexity of the decision boundary separating separable and entangled quantum states.

In the standard training scenario, $20\%$ of the dataset ($4,000$ states) was used for training, while the remaining $80\%$ ($16,000$ states) formed the test set. Due to the random train--test split, the training set contained $1,997$ separable states and $2,003$ entangled states. The resulting classification performance is summarized in Table~\ref{tab:classification_standard}.

\begin{table}[h]
\centering
\renewcommand{\arraystretch}{1.3}
\setlength{\tabcolsep}{16pt}
\caption{Classification results for the standard training scenario.}
\label{tab:classification_standard}
\begin{tabular}{lccc}
\hline
Method & Accuracy & Precision & Recall \\
\hline
Logistic Regression    & 1.0000 & 1.0000 & 1.0000 \\
KNeighborsClassifier   & 1.0000 & 1.0000 & 1.0000 \\
DecisionTreeClassifier & 0.9946 & 0.9946 & 0.9945 \\
SVC                    & 1.0000 & 1.0000 & 1.0000 \\
RandomForestClassifier & 1.0000 & 1.0000 & 1.0000 \\
\hline
\end{tabular}
\end{table}

As shown in Table~\ref{tab:classification_standard}, almost all methods achieved perfect classification performance. Only the single \texttt{DecisionTreeClassifier} showed a slight decrease in performance. The corresponding confusion matrix is presented in Table~\ref{tab:confusion_tree_standard}.

\begin{table}[h]
\centering
\renewcommand{\arraystretch}{1.3}
\setlength{\tabcolsep}{20pt}
\caption{Confusion matrix for \texttt{DecisionTreeClassifier} in the standard training scenario.}
\label{tab:confusion_tree_standard}
\begin{tabular}{lcc}
\hline
 & Predicted 0 & Predicted 1 \\
\hline
Actual 0 & 7960 & 43 \\
Actual 1 & 44 & 7953 \\
\hline
\end{tabular}
\end{table}

These results demonstrate that the SIC-POVM measurement vectors contain sufficient information to reliably distinguish between separable and maximally entangled states without explicit density-matrix reconstruction.

To further investigate the robustness of the investigated ML methods, a second scenario with an extremely limited training set was considered. In this case, only $40$ states were used for training, equally divided between the two classes ($20$ separable and $20$ entangled states). The remaining $19,960$ states were used exclusively for testing. Consequently, the training set represented only $0.2\%$ of the entire dataset. The obtained results are presented in Table~\ref{tab:classification_limited}.

\begin{table}[h]
\centering
\renewcommand{\arraystretch}{1.3}
\setlength{\tabcolsep}{16pt}
\caption{Classification results for the extremely limited training set.}
\label{tab:classification_limited}
\begin{tabular}{lccc}
\hline
Method & Accuracy & Precision & Recall \\
\hline
Logistic Regression   & 0.8487 & 0.8771 & 0.8111 \\
KNeighborsClassifier  & 1.0000 & 1.0000 & 1.0000 \\
DecisionTreeClassifier& 0.8828 & 0.9785 & 0.7828 \\
SVC                   & 1.0000 & 1.0000 & 1.0000 \\
RandomForestClassifier& 0.9980 & 0.9960 & 1.0000 \\
\hline
\end{tabular}
\end{table}

Under these challenging conditions, the best performance was achieved by \texttt{KNeighborsClassifier}, \texttt{SVC}, and \texttt{RandomForestClassifier}. This demonstrates that separable and maximally entangled states remain readily distinguishable from the $16$-dimensional measurement vectors, even when the number of available training examples is extremely small.

\begin{table*}[h]
\centering
\renewcommand{\arraystretch}{1.5}
\setlength{\tabcolsep}{10pt}
\caption{Confusion matrices for the limited training-set scenario.}
\label{tab:confusion_limited}
\begin{tabular}{lcc|cc|cc}
\hline
 & \multicolumn{2}{c|}{Logistic Regression}
 & \multicolumn{2}{c|}{DecisionTreeClassifier}
 & \multicolumn{2}{c}{RandomForestClassifier} \\
\cline{2-3}\cline{4-5}\cline{6-7}
 & Pred. 0 & Pred. 1
 & Pred. 0 & Pred. 1
 & Pred. 0 & Pred. 1 \\
\hline
Actual 0
& 8846 & 1134
& 9808 & 172
& 9940 & 40 \\

Actual 1
& 1885 & 8095
& 2168 & 7812
& 0 & 9980 \\
\hline
\end{tabular}
\end{table*}

To further examine the classification errors, confusion matrices for \texttt{Logistic Regression}, \texttt{DecisionTreeClassifier}, and \texttt{RandomForestClassifier} are presented in Table~\ref{tab:confusion_limited}. The results reveal that \texttt{Logistic Regression} and \texttt{DecisionTreeClassifier} are considerably more sensitive to the limited amount of training data than the ensemble-based \texttt{RandomForestClassifier}. In particular, \texttt{Logistic Regression} produces misclassifications in both classes, suggesting that the decision boundary is inherently nonlinear, whereas \texttt{DecisionTreeClassifier} more frequently misclassifies entangled states as separable. In contrast, \texttt{RandomForestClassifier} maintains excellent classification performance, correctly identifying all entangled states while misclassifying only a small number of separable states.

We expect that separable and maximally entangled states are well separated within the set of two-qubit quantum states, and consequently that the corresponding noisy photon-count vectors should also be readily distinguishable in the 16-dimensional SIC-POVM measurement space. This explains why most algorithms achieve very high classification performance. More significant differences between the methods appear only when the number of training examples is substantially reduced. In this case, \texttt{Logistic Regression}, as a simple linear model, and a single \texttt{DecisionTreeClassifier} are more sensitive to the reduced size of the training dataset. In contrast, \texttt{KNeighborsClassifier}, \texttt{SVC}, and \texttt{RandomForestClassifier} remain more stable, as they rely, respectively, on local relations between samples, more flexible decision boundaries, and the aggregation of multiple decision trees. These results indicate that the direct classification of noisy SIC-POVM measurement vectors is particularly robust for well-separated classes, whereas the choice of algorithm becomes more important under limited-training conditions.

Overall, these results demonstrate that information about quantum entanglement is strongly encoded in the raw SIC-POVM measurement data. The superior performance of nonlinear and ensemble-based methods further suggests that the decision boundary separating separable and entangled states is not linearly separable. Consequently, the results support the central hypothesis of this work that entanglement properties can be identified directly from noisy measurement statistics, without explicit density-matrix reconstruction or the application of conventional separability criteria.

\subsubsection{Unsupervised Clustering}

In the second stage of the analysis, unsupervised clustering was performed using the same dataset as in the supervised classification task. The dataset comprised $20,000$ noisy SIC-POVM measurement vectors, each containing $16$ coincidence counts corresponding to a two-photon polarization state. The binary class labels $\mathcal{C}^{\mathrm{bin}}$ were excluded from the clustering process and were used solely for evaluating the quality of the resulting partition.

Clustering was carried out using the \texttt{KMeans} algorithm \cite{bishop2006pattern}. As a first step, the optimal number of clusters was determined using the elbow method. For this purpose, the inertia was evaluated for values of $K$ ranging from $1$ to $8$. The resulting values are summarized in Table~\ref{tab:kmeans_inertia}.

\begin{table}[h]
\centering
\renewcommand{\arraystretch}{1.4}
\setlength{\tabcolsep}{20pt}
\caption{Inertia values for different numbers of clusters in the \texttt{KMeans} algorithm.}
\label{tab:kmeans_inertia}
\begin{tabular}{cc}
\hline
Number of clusters $K$ & Inertia \\
\hline
1 & $1.4228 \times 10^{12}$ \\
2 & $8.9178 \times 10^{11}$ \\
3 & $7.8664 \times 10^{11}$ \\
4 & $7.0596 \times 10^{11}$ \\
5 & $6.5590 \times 10^{11}$ \\
6 & $6.3421 \times 10^{11}$ \\
7 & $5.7955 \times 10^{11}$ \\
8 & $5.8316 \times 10^{11}$ \\
\hline
\end{tabular}
\end{table}

\begin{figure}[H]
\centering
\includegraphics[width=0.7\textwidth]{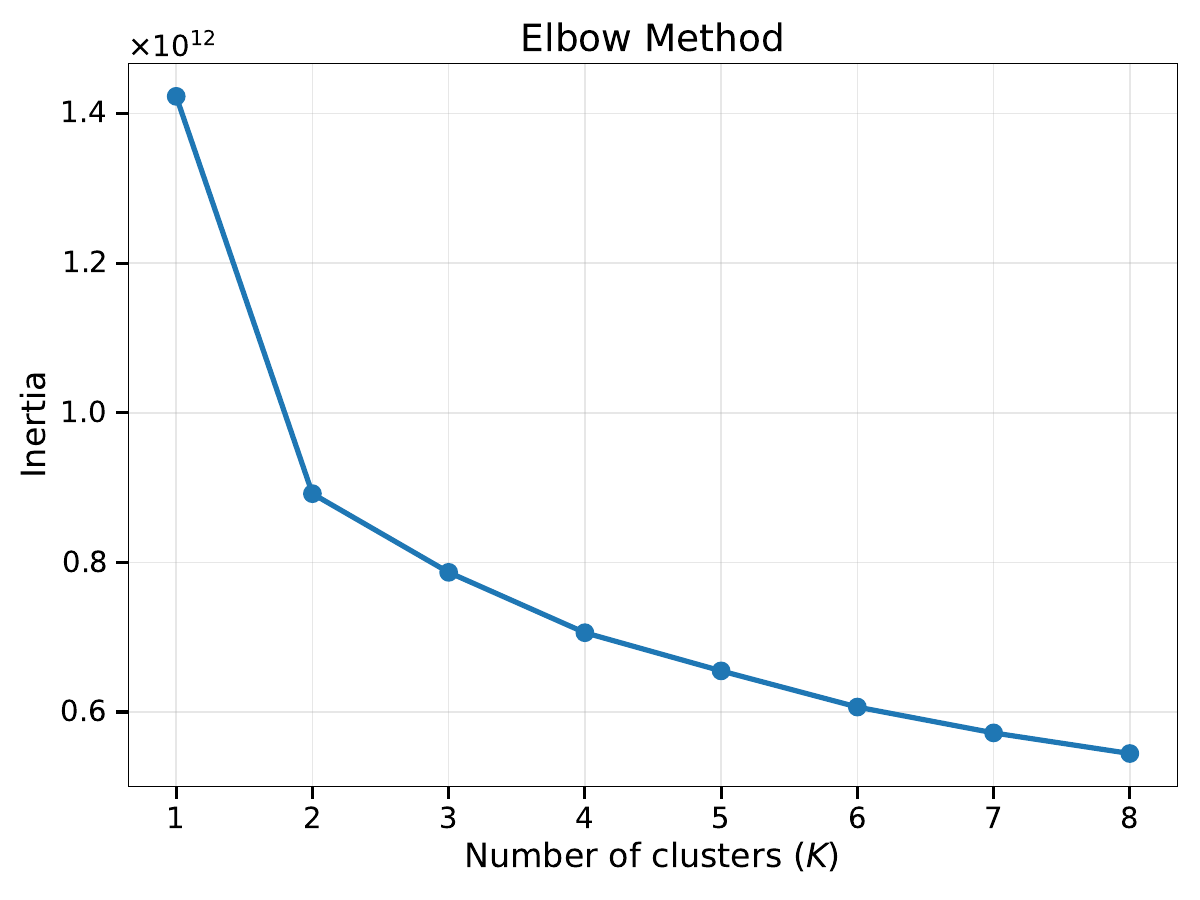}
\caption{Dependence of inertia on the number of clusters in the \texttt{KMeans} method. The largest decrease in inertia occurs for the transition from $K=1$ to $K=2$, which indicates two natural groups in the measurement-data space.}
\label{fig:elbow_method}
\end{figure}

The largest decrease in inertia occurs when going from $K=1$ to $K=2$, which can be observed in Figure~\ref{fig:elbow_method}. Further increasing the number of clusters leads to much smaller changes. Therefore, the elbow method indicates that the optimal number of clusters is
\begin{equation}
K_{\mathrm{opt}}=2.
\end{equation}

After choosing $K=2$, clustering of the entire dataset was performed. The algorithm divided the data into two almost equally numerous clusters, as shown in Table~\ref{tab:kmeans_cluster_sizes}.

\begin{table}[H]
\centering
\renewcommand{\arraystretch}{1.3}
\setlength{\tabcolsep}{20pt}
\caption{Number of observations assigned to individual clusters by the \texttt{KMeans} algorithm.}
\label{tab:kmeans_cluster_sizes}
\begin{tabular}{cc}
\hline
Cluster & Number of observations \\
\hline
0 & 10,005 \\
1 & 9,995 \\
\hline
\end{tabular}
\end{table}

Since cluster labels in unsupervised learning have no predefined physical meaning, the correspondence between cluster indices and the true class labels was determined after clustering. Following this relabeling, the comparison matrix shown in Table~\ref{tab:kmeans_confusion} was obtained.

\begin{table}[H]
\centering
\renewcommand{\arraystretch}{1.3}
\setlength{\tabcolsep}{20pt}
\caption{Comparison matrix of \texttt{KMeans} clustering results with the true labels after correction of cluster numbering.}
\label{tab:kmeans_confusion}
\begin{tabular}{lcc}
\hline
 & Cluster assigned as 0 & Cluster assigned as 1 \\
\hline
Actual 0 & 9,995 & 5 \\
Actual 1 & 0 & 10,000 \\
\hline
\end{tabular}
\end{table}

The clustering results demonstrate that \texttt{KMeans}, despite having no access to the class labels during training, almost perfectly recovered the natural partition of the dataset into separable and maximally entangled states. Out of $20,000$ observations, only five were assigned to the incorrect cluster. All misclassified samples corresponded to separable states that were grouped together with the entangled states. The resulting clustering accuracy is
\begin{equation}
\mathrm{accuracy}
=
\frac{9995+10000}{20000}
=
0.99975.
\end{equation}

An interesting feature of the clustering results is that all entangled states were assigned to the same cluster, whereas the only misclassified observations correspond to separable states. Since the \texttt{KMeans} algorithm had no access to the class labels during clustering, this asymmetry reflects the intrinsic structure of the measurement data rather than knowledge acquired during supervised learning. The results suggest that, in the sixteen-dimensional space of noisy SIC-POVM measurement vectors, the subset of entangled states forms a well-defined region that is naturally identified by the clustering algorithm. In contrast, the measurement vectors corresponding to separable states appear to occupy a more broadly distributed region, with a small number of observations located close to the boundary separating the two groups. After the inclusion of Poissonian shot noise, these boundary cases become indistinguishable from nearby entangled states, leading to the five observed misclassifications.

\subsection{Werner States}\label{sec:Werner States}

This section presents the analysis of the Werner-state dataset. In contrast to the previous dataset, which included separable and maximally entangled states, Werner states make it possible to study cases with varying degrees of entanglement. This allows not only the classification of separable and entangled states, but also the estimation of a continuous entanglement measure described by concurrence. The following part presents the results of classification and nonlinear regression for this dataset.

Werner states were selected because they constitute one of the most widely studied families of mixed two-qubit states and provide a convenient benchmark for investigating entanglement. Their simple one-parameter structure allows the degree of entanglement to vary continuously from fully separable states to maximally entangled Bell states, thereby covering the entire range of concurrence values. This makes them particularly well suited for systematically evaluating both binary classification and regression methods for entanglement quantification.

It should be emphasized that the objective of this work was not to investigate the geometric distribution of quantum states in the Hilbert space or to generate state ensembles according to a particular physical measure. Instead, our goal was to construct a diverse and controlled dataset for comparing different ML approaches across the full range of entanglement. To avoid restricting the analysis to the highly symmetric one-parameter manifold of Werner states, independent random local unitary transformations were applied to every generated state. Since local unitary operations preserve concurrence, this procedure produced a large ensemble of geometrically distinct quantum states while maintaining the same degree of entanglement. Consequently, for each concurrence value, the dataset contains a broad collection of physically different quantum states rather than a single representative of the Werner family, providing a substantially richer benchmark for the analysis of ML methods.

\subsubsection{Supervised Classification of Werner States}\label{sec:werner_classification}

The classification task was subsequently extended from the dataset of maximally entangled and separable states to the considerably larger dataset generated from the family of Werner states. This dataset contained $250,000$ observations corresponding to noisy SIC-POVM measurement vectors obtained from two-qubit quantum states exhibiting different degrees of entanglement as described in Section~\ref{IIIB}.

The dataset was not perfectly balanced with respect to the binary entanglement label. It contained $167,002$ entangled states and $82,998$ separable states. In this classification problem, the entangled-state class $\mathcal{C}^{\mathrm{bin}}=1$ was defined as the set of Werner states satisfying $p>1/3$, while states with $p\leq 1/3$ were assigned to the separable-state class $\mathcal{C}^{\mathrm{bin}}=0$.

Prior to the ML analysis, the dataset was examined for duplicate observations. No duplicated rows were identified, indicating that all generated measurement vectors were unique. Consequently, the entire dataset containing $250,000$ observations was retained for further analysis. A standard train-test split was employed, with $80\%$ of the data used for training and the remaining $20\%$ ($50,000$ observations) reserved for testing.

The binary classification task consisted of assigning each measurement vector to either the separable-state class or the entangled-state class. Four ML algorithms were investigated: \texttt{KNeighborsClassifier}, \texttt{DecisionTreeClassifier}, \texttt{SVC}, and \texttt{RandomForestClassifier}. It is worth emphasizing that, in the case of the Werner-state analysis, \texttt{Logistic Regression} was excluded because, under class imbalance conditions, it tended to predict the majority class (i.e. entangled class), without providing a meaningful separation between the classes. The resulting classification metrics are summarized in Table~\ref{tab:werner_metrics}.

\begin{table}[H]
\centering
\renewcommand{\arraystretch}{1.3}
\setlength{\tabcolsep}{16pt}
\caption{Classification performance for the Werner-state dataset.}
\label{tab:werner_metrics}
\begin{tabular}{lcccc}
\hline
Method &
Accuracy &
Precision &
Recall &
Misclassifications \\
\hline
KNeighborsClassifier   & 0.98940 & 0.99378 & 0.99033 & 530 \\
DecisionTreeClassifier & 0.97902 & 0.98726 & 0.98126 & 1049 \\
RandomForestClassifier & 0.99132 & 0.99549 & 0.99150 & 434 \\
SVC                    & 0.99498 & 0.99647 & 0.99602 & 251 \\
\hline
\end{tabular}
\end{table}

The confusion matrices obtained for the \texttt{KNeighborsClassifier}, \texttt{DecisionTreeClassifier}, \texttt{RandomForestClassifier}, and \texttt{SVC} models are presented in Table~\ref{tab:cm_werner}.

\begin{table}[H]
\centering
\renewcommand{\arraystretch}{1.5}
\setlength{\tabcolsep}{18pt}
\caption{Confusion matrices for the supervised classification of Werner states.}
\label{tab:cm_werner}
\begin{tabular}{lcc}
\hline
 & Predicted 0 & Predicted 1 \\
\hline

\multicolumn{3}{c}{\texttt{KNeighborsClassifier}}\\
Actual 0 & 16393 & 207 \\
Actual 1 & 323 & 33077 \\
\hline

\multicolumn{3}{c}{\texttt{DecisionTreeClassifier}}\\
Actual 0 & 16177 & 423 \\
Actual 1 & 626 & 32774 \\
\hline

\multicolumn{3}{c}{\texttt{RandomForestClassifier}}\\
Actual 0 & 16450 & 150 \\
Actual 1 & 284 & 33116 \\
\hline

\multicolumn{3}{c}{\texttt{SVC}}\\
Actual 0 & 16482 & 118 \\
Actual 1 & 133 & 33267 \\
\hline
\end{tabular}
\end{table}

As shown in Table~\ref{tab:werner_metrics}, all analyzed models achieved high classification performance for Werner states. The best result was obtained by the \texttt{SVC} model, for which the accuracy was $0.99498$, and the number of misclassifications was the lowest among all investigated methods. Very high performance was also achieved by the \texttt{RandomForestClassifier} model, with an accuracy of $0.99132$. Slightly lower, but still high, results were obtained for \texttt{KNeighborsClassifier} and \texttt{DecisionTreeClassifier}.

To analyze the character of the errors in more detail, the confusion matrices for the four investigated models are presented in Table~\ref{tab:cm_werner}. These results show that all models correctly identified the vast majority of separable and entangled states. The largest number of errors was made by the single \texttt{DecisionTreeClassifier} model, which indicates its greater sensitivity to the complex structure of the measurement data. The smallest number of misclassifications was obtained by the \texttt{SVC} model, which best reproduced the boundary separating separable and entangled states in the Werner-state dataset.

Compared with the dataset containing only separable and maximally entangled states, the classification of Werner states represents a more difficult task, because it includes states with varying degrees of entanglement as well as cases located closer to the separability boundary. Nevertheless, the obtained results indicate that noisy SIC-POVM measurement vectors still contain sufficient information to effectively distinguish separable and entangled states.

We observed that, for all investigated ML methods, the majority of misclassifications correspond to entangled states that were incorrectly classified as separable. We hypothesized that this behavior originates from the properties of Werner states close to the separability threshold, where the amount of entanglement is very small. In the presence of measurement noise, such weakly entangled states are expected to produce SIC-POVM measurement vectors that closely resemble those of separable states, making them more difficult to classify correctly.

To verify this hypothesis, in addition to the standard classification metrics, information about all misclassified states was collected. For each erroneous prediction, the corresponding value of the Werner-state parameter $p$ was recorded. This enabled the distribution of classification errors to be analyzed as a function of the Werner-state parameter, providing insight into how the misclassifications are related to the transition between separable and entangled states.

\begin{figure}[H]
\centering

\begin{minipage}{0.49\textwidth}
\centering
\includegraphics[width=\linewidth]{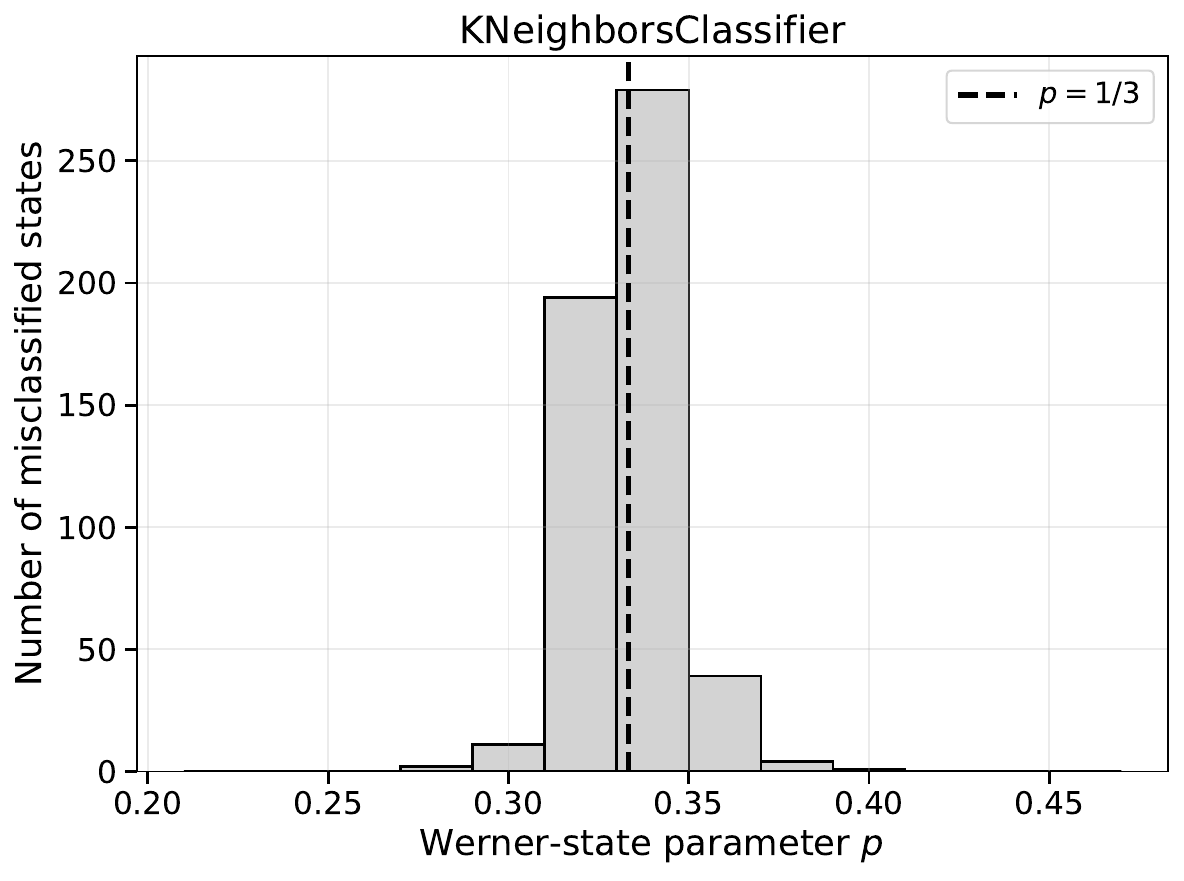}
\put(-250,0){(a)}
\end{minipage}
\hfill
\begin{minipage}{0.49\textwidth}
\centering
\includegraphics[width=\linewidth]{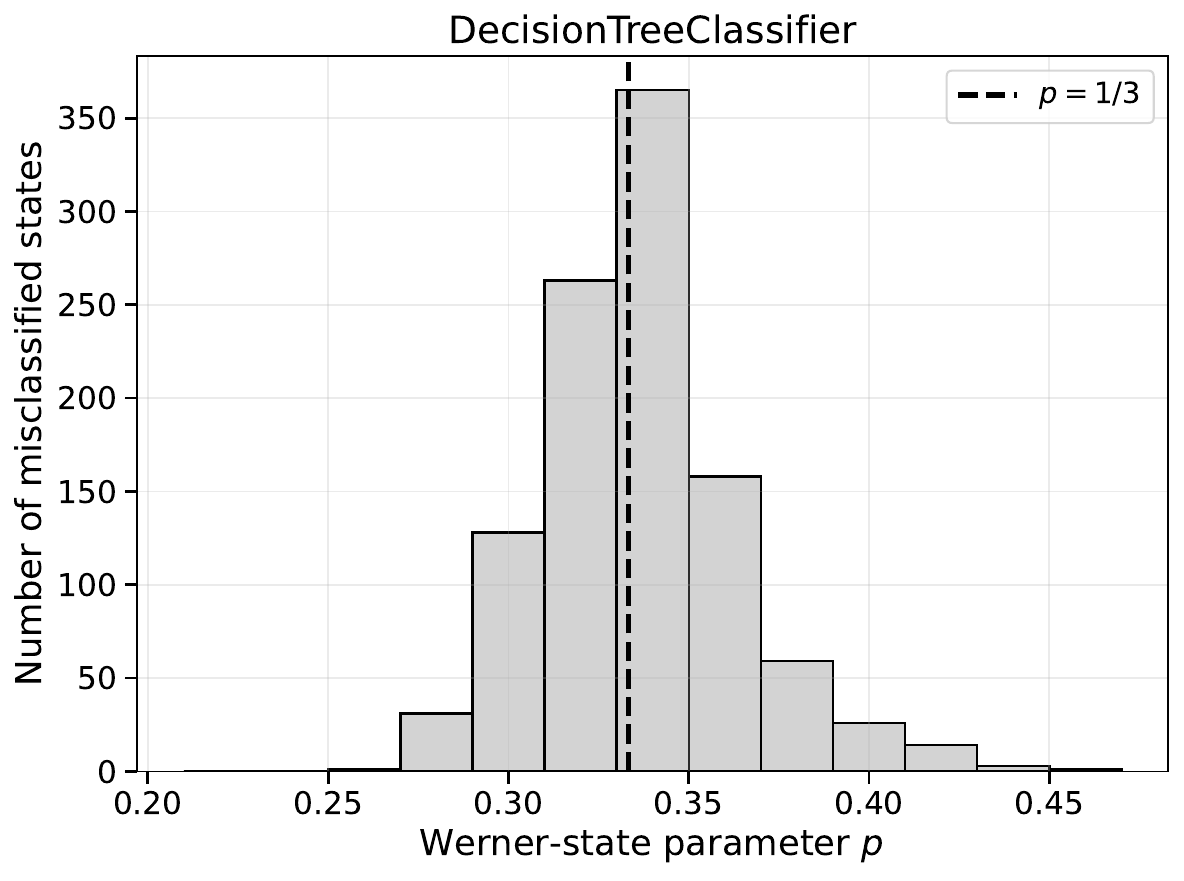}
\put(-250,0){(b)}
\end{minipage}
\vspace{0.3cm}
\begin{minipage}{0.49\textwidth}
\centering
\includegraphics[width=\linewidth]{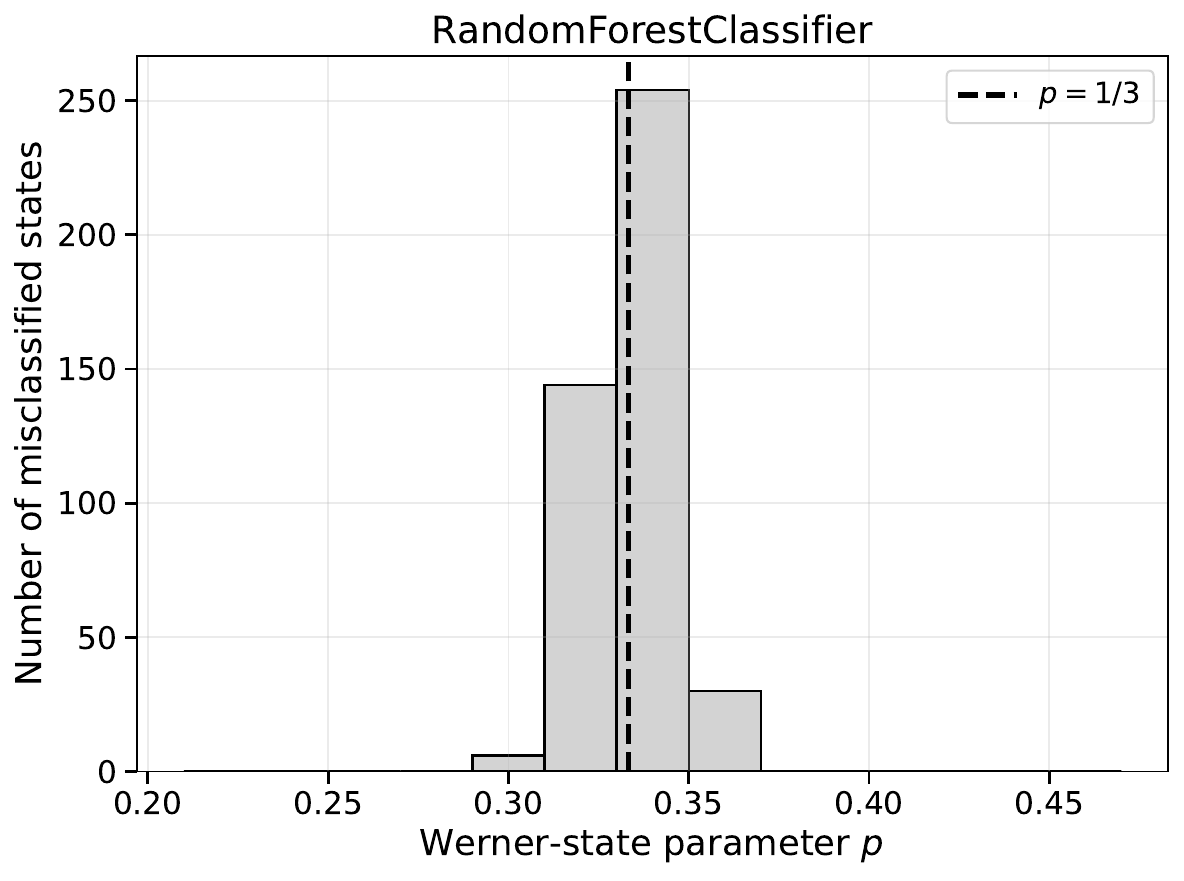}
\put(-250,0){(c)}
\end{minipage}
\hfill
\begin{minipage}{0.49\textwidth}
\centering
\includegraphics[width=\linewidth]{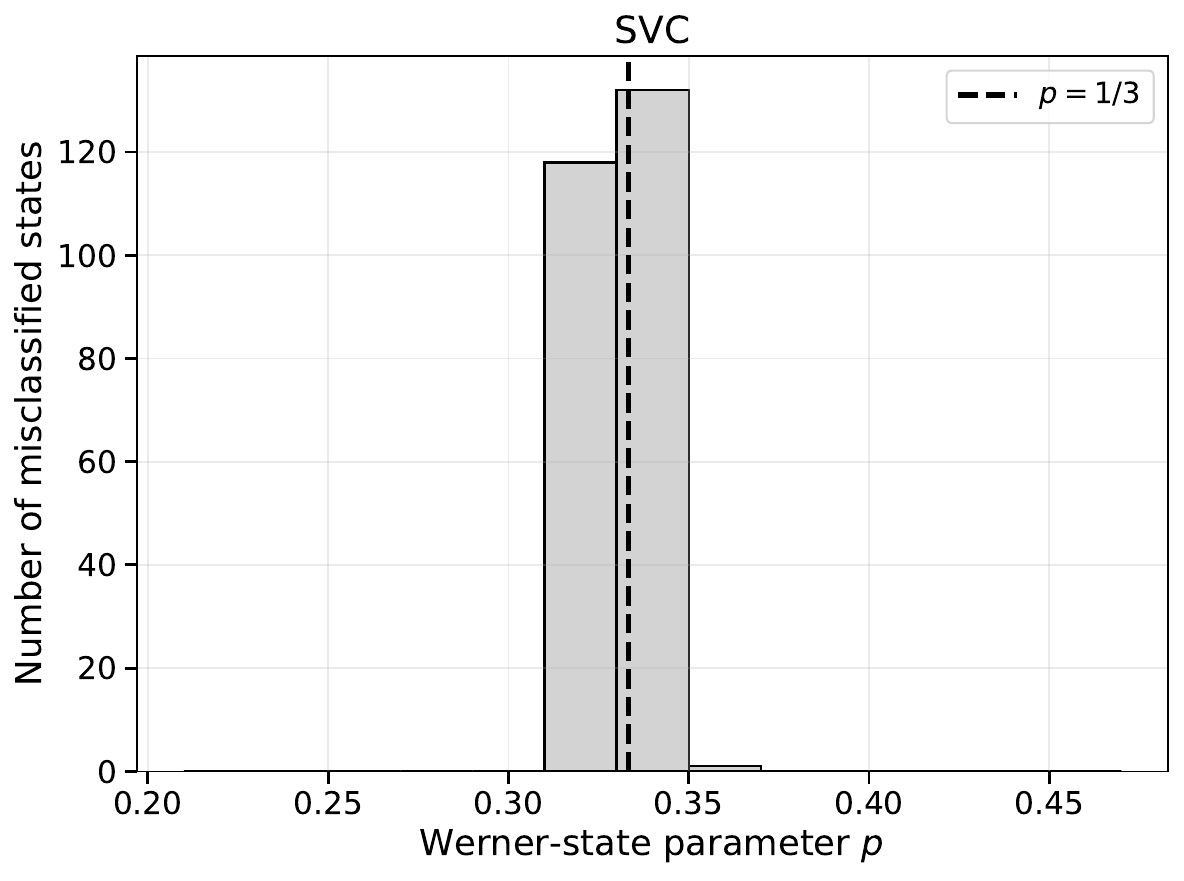}
\put(-250,0){(d)}
\end{minipage}
\caption{Histograms showing the distribution of misclassified states as a function of the Werner-state parameter $p$ for (a) \texttt{KNeighborsClassifier}, (b) \texttt{DecisionTreeClassifier}, (c) \texttt{RandomForestClassifier}, and (d) \texttt{SVC}.}
\label{fig:error_histograms_werner}
\end{figure}

Figure~\ref{fig:error_histograms_werner} presents histograms of classification errors as a function of the Werner state parameter $p$ for the four investigated ML models. The distributions of misclassified states show that the errors of all analyzed models are primarily concentrated near the separability boundary at $p = 1/3$. This indicates that the most difficult states to classify are those located close to the transition between the separable and entangled regions. For parameter values that are clearly distant from this boundary, the number of errors decreases rapidly, which suggests that the models are significantly more confident in recognizing states that are either clearly separable or strongly entangled.

Among the analyzed methods, the largest spread of errors is observed for the \texttt{DecisionTreeClassifier}, which is consistent with the highest number of misclassifications produced by this model. Both \texttt{KNeighborsClassifier} and \texttt{RandomForestClassifier} exhibit a narrower error distribution concentrated around $p = 1/3$. The most concentrated error distribution is obtained for the \texttt{SVC} model, for which misclassifications occur almost exclusively in the immediate vicinity of the separability boundary. This result confirms that the \texttt{SVC} model best reproduces the decision boundary separating separable and entangled states in the analyzed Werner-state dataset.

\subsubsection{Nonlinear Regression}

In the final stage of the analysis, the problem of quantitative estimation of quantum entanglement using nonlinear regression methods was considered \cite{bishop2006pattern}. In contrast to the previous classification task, in which separable and entangled states were distinguished for two-qubit systems, the aim here was to predict a continuous measure describing the degree of entanglement. As the target variable, concurrence $C$ was selected, taking values in the interval

\begin{equation}
0 \leq C \leq 1.
\end{equation}

The same dataset that had previously been used in the supervised classification of Werner states (Section~\ref{sec:werner_classification}) was employed in the present study. The input data consisted of 16-dimensional vectors of pseudo-experimental SIC-POVM counts affected by Poisson noise, whereas the dependent variable corresponded to the numerically computed concurrence value for the corresponding quantum state, $y=C$. For comparison, two ML methods were investigated for this task: \texttt{Support Vector Regression} (SVR) and \texttt{Random Forest Regressor} (RFR).

Initially, the prepared regression dataset contained $250,000$ observations. However, the full dataset turned out to be computationally too large for the classical implementation of \texttt{SVR} with an RBF kernel, because the complexity of this method grows rapidly with the number of training samples. Therefore, a randomly selected representative sample of $5,000$ measurement vectors was used for the actual computations. Before training, the input data were standardized using \texttt{StandardScaler} \cite{Pedregosa2011}, and then an \texttt{SVR} model with an RBF kernel was applied, using a regularization parameter equal to 10 and an $\epsilon$-insensitive loss parameter of $\epsilon = 0.01$.

\begin{table}[H]
\centering
\renewcommand{\arraystretch}{1.5}
\setlength{\tabcolsep}{19pt}
\caption{Regression performance for concurrence estimation using quantum state tomography (QST), Support Vector Regression (SVR), and Random Forest Regressor (RFR).}
\label{tab:regression_results}
\begin{tabular}{lccccc}
\hline
 & QST & \multicolumn{2}{c}{SVR} & \multicolumn{2}{c}{RFR} \\
\cline{2-2}\cline{3-4}\cline{5-6}
Sample size & 5,000 & \multicolumn{2}{c}{5,000} & \multicolumn{2}{c}{250,000} \\
\hline
Metric & ALL & TRAIN & TEST & TRAIN & TEST \\
\hline
MAE   & 0.0066 & 0.00748 & 0.00865 & 0.00137 & 0.00315 \\
MSE   & 0.00014 & 0.00011 & 0.00015 & 0.00001 & 0.00005 \\
$R^2$ & 0.99882 & 0.99876 & 0.99829 & 0.99990 & 0.99947 \\
\hline
\end{tabular}
\end{table}

As an alternative nonlinear regression approach, a \texttt{RFR} model was also investigated. In contrast to \texttt{SVR}, tree-based ensemble methods exhibit considerably better computational scalability with increasing dataset size. Therefore, the RFR model was trained using the complete dataset consisting of $250,000$ observations. The adopted model comprised $200$ decision trees, while all remaining hyperparameters were set to their standard values as implemented in the \texttt{scikit-learn} library \cite{Pedregosa2011}.

For both methods, the test set was obtained by randomly selecting $20\%$ of the corresponding dataset. The prediction quality was evaluated using the mean absolute error (MAE), mean squared error (MSE), and the coefficient of determination $R^2$. Both investigated regression methods achieved excellent predictive performance, as summarized in Table~\ref{tab:regression_results}.

The \texttt{SVR} model yielded a coefficient of determination of $R^2=0.99829$ for the test set together with a mean absolute error of $\mathrm{MAE}=0.00865$, corresponding to an average concurrence estimation error below $1\%$. The relatively small difference between the training and test metrics indicates good generalization performance without clear evidence of overfitting.

The RFR further improved the regression accuracy. When trained on the complete dataset, it achieved $R^2=0.99947$ on the test set while reducing the mean absolute error to $\mathrm{MAE}=0.00315$. The corresponding training metrics remained similarly close to the test results, indicating stable generalization despite the increased model complexity. These results demonstrate that both nonlinear regression approaches are capable of accurately estimating concurrence directly from noisy SIC-POVM measurement statistics, while the ensemble-based RFR model provides the highest predictive accuracy among the investigated methods.

To benchmark the proposed ML-based regression methods against a conventional approach, an additional analysis based on QST was performed. The same subset of $5,000$ noisy SIC-POVM measurement vectors that had been used for the SVR model was utilized for density matrix reconstruction using the standard least-squares estimator. Subsequently, the concurrence of each reconstructed density matrix was evaluated analytically according to its standard definition. The reconstructed concurrence values were then compared with the corresponding reference values, allowing the same performance metrics (MAE, MSE, and $R^2$) to be computed as for the ML-based regression models. The resulting comparison is summarized in Table~\ref{tab:regression_results}.

The obtained results show that the conventional QST approach achieves predictive accuracy comparable to that of the SVR model. Both methods yield very similar values of MAE, MSE, and the coefficient of determination, indicating that the SVR model is capable of reproducing concurrence estimates with an accuracy comparable to that obtained through full density matrix reconstruction. In contrast, the RFR substantially improves the estimation quality, reducing both the MAE and MSE while increasing the coefficient of determination. Moreover, unlike the QST-based approach, both ML methods bypass the computationally demanding process of density matrix reconstruction and the subsequent analytical evaluation of concurrence. This makes the ML framework particularly attractive for the efficient analysis of large sets of experimental data, with the RFR model offering the best overall trade-off between computational efficiency and estimation accuracy among the investigated approaches.

\subsection{Discussion}

The obtained results in Sections \ref{sec:Separable and Maximally Entangled States}, \ref{sec:Werner States} indicate that SIC-POVM measurement statistics preserve information about entanglement in a manner that can be exploited by ML algorithms. It is important that the models operated directly on noisy photon-count vectors rather than reconstructed density matrices, which implies that entanglement properties can be inferred from measurement data at an earlier stage of processing than in standard quantum state tomography.

The classification results show that algorithms such as \texttt{KNeighborsClassifier}, \texttt{SVC}, and \texttt{RandomForestClassifier} achieve practically perfect performance in distinguishing separable and maximally entangled states. Particularly interesting is the behavior of the models under an extreme limitation of the training set to only $40$ examples, where despite such a small number of training samples all selected algorithms maintain an accuracy of $100\%$, suggesting the existence of a very well-defined geometric structure of the SIC-POVM measurement space. The obtained results further suggest that the decision boundary in this space is nonlinear in nature, as evidenced by the high performance of models capable of capturing complex decision structures, alongside the reduced effectiveness of linear approaches. An additional indication of the nonlinear nature of the decision boundary is the sensitivity of logistic regression to class imbalance, which, in the analysis of Werner states, leads to a dominance of majority-class predictions.

In parallel, the analysis of the classification error distributions as a function of the Werner state parameter $p$ indicates that the difficulty of discrimination is not uniformly distributed across the parameter space but is concentrated near the separability boundary at $p=1/3$. In this transitional region, the geometric structure of the SIC-POVM measurement space becomes most “blurred” from the perspective of ML models, leading to increased ambiguity in classification decisions. Conversely, the rapid decrease in the number of errors with increasing distance from this boundary indicates that, for clearly separable and strongly entangled states, the separation between classes becomes increasingly pronounced, enabling stable performance of all analyzed models.

In the case of supervised classification and unsupervised clustering of the dataset containing separable and maximally entangled states, the results are consistent, also suggesting the existence of a well-defined geometric structure in the 16-dimensional SIC-POVM measurement space. Since clustering was performed without using class labels, the obtained separation can be interpreted as a natural partition arising from the physical properties of the data. In particular, the elbow method indicates that the optimal number of clusters is $K=2$, which is consistent with the physical division of states into separable and entangled ones, while the \texttt{KMeans} algorithm almost perfectly reproduces the true data partition, making only a few assignment errors. This result confirms that the analyzed dataset possesses a natural geometric structure corresponding to the physical properties of quantum states, which can be detected using unsupervised learning methods without using labels. The results obtained for Werner states further show that this structure remains stable even in a more complex scenario involving a continuous range of entanglement values. The error distribution analysis indicates that the difficulty of classification is concentrated near the separability boundary at $p=1/3$, which is consistent with the expected transition between separable and entangled states.

The regression results show that the same measurement representation can be used not only for classification but also for predicting a continuous measure of entanglement. The \texttt{SVR} model with an \texttt{RBF} kernel achieves very high predictive accuracy, obtaining a coefficient of determination $R^2 \approx 0.9983$ on the test data. The second model implemented in this study, \texttt{RFR}, further improves the prediction quality, achieving $R^2 \approx 0.9995$ while simultaneously reducing both the MAE and the MSE. Moreover, the accuracy of \texttt{RFR} is significantly better than the performance of QST-based estimation of concurrence. This demonstrates that scalable ensemble-based regression methods are capable of exploiting the full dataset and provide more accurate concurrence estimation than the computationally more demanding \texttt{SVR} approach. These results have practical significance, since after training the regression models can operate without full state reconstruction and without explicit computation of concurrence from the density matrix. Such approaches may therefore serve as fast preliminary tools for analyzing large experimental datasets.

The noise model considered in this study is limited to Poissonian fluctuations associated with finite photon-counting statistics. Although this model captures an important source of statistical uncertainty in optical measurements, it does not account for additional experimental imperfections, such as channel losses, dark counts, detector dead time, or multi-photon emission events. These effects may introduce systematic biases and modify the statistical distribution of the measurement vectors in ways that are not represented by the present simulations. Consequently, the reported performance should be interpreted as applying to data affected by Poissonian counting noise rather than to fully realistic experimental conditions. Further validation using more comprehensive noise models and experimental measurement data is therefore required.

In summary, the obtained results confirm that ML methods can serve as an effective tool for supporting the analysis of tomographic data in quantum optics. In parallel, they should be regarded as a numerical validation of the concept rather than a final confirmation of its universality. A natural direction for further research is to extend the analysis to real experimental data and to compare the performance of the models with classical procedures for density matrix reconstruction and separability testing.

\section{Conclusions}\label{conclusions}

In this work, we investigated the application of ML methods to the analysis of two-qubit quantum states describing polarization-entangled photon pairs. The analysis was based on pseudo-experimental measurement data generated using SIC-POVM operators. The generated data modelled Poissonian photon-counting noise under controlled simulated conditions. Each quantum state was represented by a $16$-dimensional count vector corresponding to the measurement operators.

This work shows that SIC-POVM noisy measurement statistics contain sufficient information to enable both classification and quantitative estimation of entanglement using ML methods, without the need for density matrix reconstruction. The obtained results indicate that the structure associated with quantum entanglement is directly encoded in the measurement space and can be effectively exploited by both supervised and unsupervised learning methods. These findings demonstrate that ML provides a promising framework for the rapid analysis of quantum-optical measurement data and may serve as an efficient alternative to conventional state-reconstruction-based workflows in applications where fast entanglement characterization is required.

The comparison with conventional QST further demonstrates that ML-based regression can provide concurrence estimates of comparable or even higher accuracy while avoiding the computational overhead associated with full density matrix reconstruction. In particular, the \texttt{RFR} outperformed the tomographic baseline in terms of both prediction accuracy and computational efficiency, illustrating the potential of scalable ensemble-based methods for the rapid analysis of large quantum datasets. Although the present study is limited to simulated measurement data affected by Poissonian counting noise, the obtained results establish a proof of concept and motivate future validation using more comprehensive experimental noise models and real quantum-optical measurements.

\section*{Funding}
Not applicable.

\section*{Data Availability}
The raw data supporting the conclusions of this article will be made available by the corresponding author on request.

\section*{Conflict of Interest}
The authors declare that there is no conflict of interest.

\bibliographystyle{unsrt} \bibliography{references}

@book{Nielsen2000,
  author    = {Michael A. Nielsen and Isaac L. Chuang},
  title     = {Quantum Computation and Quantum Information},
  publisher = {Cambridge University Press},
  year      = {2000},
  address   = {Cambridge}
}

@article{james2001measurement,
  title={Measurement of qubits},
  author={James, Daniel FV and Kwiat, Paul G and Munro, William J and White, Andrew G},
  journal={Phys. Rev. A},
  volume={64},
  number={5},
  pages={052312},
  year={2001},
  publisher={APS},
  doi={10.1103/PhysRevA.64.052312}
}

@article{vrehavcek2004minimal,
  title={Minimal qubit tomography},
  author={{\v{R}}eh{\'a}{\v{c}}ek, Jaroslav and Englert, Berthold-Georg and Kaszlikowski, Dagomir},
  journal={Phys. Rev. A},
  volume={70},
  number={5},
  pages={052321},
  year={2004},
  publisher={APS},
  doi={10.1103/PhysRevA.70.052321}
}

@article{Renes2004,
  title={Symmetric informationally complete quantum measurements},
  author={Renes, Joseph M and Blume-Kohout, Robin and Scott, Andrew J and Caves, Carlton M},
  journal={J. Math. Phys.},
  volume={45},
  number={6},
  pages={2171--2180},
  year={2004},
  publisher={American Institute of Physics},
  doi={10.1063/1.1737053}
}

@article{Fuchs2017,
  title={The SIC question: History and state of play},
  author={Fuchs, Christopher A and Hoang, Michael C and Stacey, Blake C},
  journal={Axioms},
  volume={6},
  number={3},
  pages={21},
  year={2017},
  publisher={MDPI},
  doi={10.3390/axioms6030021}
}

@article{sedziak2020tomography,
  title={Tomography of time-bin quantum states using time-resolved detection},
  author={Sedziak-Kacprowicz, Karolina and Czerwinski, Artur and Kolenderski, Piotr},
  journal={Phys. Rev. A},
  volume={102},
  number={5},
  pages={052420},
  year={2020},
  publisher={APS},
  doi={10.1103/PhysRevA.102.052420}
}

@article{hill1997entanglement,
  title={Entanglement of a pair of quantum bits},
  author={Hill, Sam A and Wootters, William K},
  journal={Phys. Rev. Lett.},
  volume={78},
  number={26},
  pages={5022},
  year={1997},
  publisher={APS},
  doi={doi:10.1103/PhysRevLett.78.5022}
}

@book{blum2012density,
  title={Density matrix theory and applications},
  author={Blum, Karl},
  volume={64},
  year={2012},
  publisher={Springer Science \& Business Media}
}

@book{bishop2006pattern,
  title={Pattern Recognition and Machine Learning},
  author={Bishop, Christopher M.},
  year={2006},
  publisher={Springer}
}

@article{Horodecki2009,
  title = {Quantum entanglement},
  author = {Horodecki, Ryszard and Horodecki, Pawe\l{} and Horodecki, Micha\l{} and Horodecki, Karol},
  journal = {Rev. Mod. Phys.},
  volume = {81},
  issue = {2},
  pages = {865--942},
  numpages = {0},
  year = {2009},
  month = {Jun},
  publisher = {American Physical Society},
  doi = {10.1103/RevModPhys.81.865},
  url = {https://link.aps.org/doi/10.1103/RevModPhys.81.865}
}

@article{Gisin2007,
  title={Quantum communication},
  author={Gisin, Nicolas and Thew, Rob},
  journal={Nat. Photonics},
  volume={1},
  number={3},
  pages={165--171},
  year={2007},
  doi={10.1038/nphoton.2007.22}
}

@article{Pirandola2020,
  title={Advances in quantum cryptography},
  author={Pirandola, Stefano and Andersen, Ulrik L. and Banchi, Leonardo and Berta, Mario and Bunandar, Darius and Colbeck, Roger and Englund, Dirk and Gehring, Tobias and Lupo, Cosmo and Ottaviani, Carlo and Pereira, Joseph and Razavi, Mohsen and Shaari, Jonatan S. and Tomamichel, Marco and Usenko, Vladyslav C. and Vallone, Giuseppe and Villoresi, Paolo and Wallden, Petros},
  journal={Adv. Opt. Photonics.},
  volume={12},
  number={4},
  pages={1012--1236},
  year={2020},
  doi={10.1364/AOP.361502}
}

@article{Peres1996,
  author  = {Peres, Asher},
  title   = {Separability Criterion for Density Matrices},
  journal = {Phys. Rev. Lett.},
  volume  = {77},
  number  = {8},
  pages   = {1413--1415},
  year    = {1996},
  doi     = {10.1103/PhysRevLett.77.1413}
}

@article{Horodecki1996,
  author  = {Horodecki, Micha{\l} and Horodecki, Pawe{\l} and Horodecki, Ryszard},
  title   = {Separability of Mixed States: Necessary and Sufficient Conditions},
  journal = {Phys. Lett. A},
  volume  = {223},
  number  = {1--2},
  pages   = {1--8},
  year    = {1996},
  doi     = {10.1016/S0375-9601(96)00706-2}
}

@article{Wootters1998,
  author  = {Wootters, William K.},
  title   = {Entanglement of Formation of an Arbitrary State of Two Qubits},
  journal = {Phys. Rev. Lett.},
  volume  = {80},
  number  = {10},
  pages   = {2245--2248},
  year    = {1998},
  doi     = {10.1103/PhysRevLett.80.2245}
}

@article{Carleo2019,
  author  = {Carleo, Giuseppe and Cirac, Ignacio and Cranmer, Kyle and Daudet, Laurent and Schuld, Maria and Tishby, Naftali and Vogt-Maranto, Leslie and Zdeborov{\'a}, Lenka},
  title   = {Machine learning and the physical sciences},
  journal = {Rev. Mod. Phys.},
  volume  = {91},
  number  = {4},
  pages   = {045002},
  year    = {2019},
  doi     = {10.1103/RevModPhys.91.045002}
}

@article{Biamonte2017,
  author  = {Biamonte, Jacob and Wittek, Peter and Pancotti, Nicola and Rebentrost, Patrick and Wiebe, Nathan and Lloyd, Seth},
  title   = {Quantum machine learning},
  journal = {Nature},
  volume  = {549},
  number  = {7671},
  pages   = {195--202},
  year    = {2017},
  doi     = {10.1038/nature23474}
}

@article{urena2024entanglement,
  title={Entanglement detection with classical deep neural networks},
  author={Ure{\~n}a, Julio and Sojo, Antonio and Bermejo-Vega, Juani and Manzano, Daniel},
  journal={Sci. Rep.},
  volume={14},
  number={1},
  pages={18109},
  year={2024},
  publisher={Nature Publishing Group UK London},
  doi     = {https://doi.org/10.1038/s41598-024-68213-0}
}

@article{asif2023entanglement,
  title={Entanglement detection with artificial neural networks},
  author={Asif, Naema and Khalid, Uman and Khan, Awais and Duong, Trung Q and Shin, Hyundong},
  journal={Sci. Rep.},
  volume={13},
  number={1},
  pages={1562},
  year={2023},
  publisher={Nature Publishing Group UK London},
  doi = {https://doi.org/10.1038/s41598-023-28745-3}
}

@article{huang2022measuring,
  title={Measuring quantum entanglement from local information by machine learning},
  author={Huang, Yulei and Che, Liangyu and Wei, Chao and Xu, Feng and Nie, Xinfang and Li, Jun and Lu, Dawei and Xin, Tao},
  journal={arXiv preprint arXiv:2209.08501},
  year={2022},
  doi = {https://doi.org/10.48550/arXiv.2209.08501}
}

@article{koutny2023deep,
  title={Deep learning of quantum entanglement from incomplete measurements},
  author={Koutn{\`y}, Dominik and Gin{\'e}s, Laia and Mocza{\l}a-Dusanowska, Magdalena and H{\"o}fling, Sven and Schneider, Christian and Predojevi{\'c}, Ana and Je{\v{z}}ek, Miroslav},
  journal={Sci. Adv.},
  volume={9},
  number={29},
  pages={eadd7131},
  year={2023},
  publisher={American Association for the Advancement of Science},
  doi = {https://doi.org/10.1126/sciadv.add7131}
}

@article{ma2024neural,
  title={Neural networks for quantum state tomography with constrained measurements},
  author={Ma, Hailan and Dong, Daoyi and Petersen, Ian R and Huang, Chang-Jiang and Xiang, Guo-Yong},
  journal={Quantum Inf. Process.},
  volume={23},
  number={9},
  pages={317},
  year={2024},
  publisher={Springer},
  doi = {https://doi.org/10.1007/s11128-024-04522-7}
}

@article{Pedregosa2011,
  title   = {Scikit-learn: Machine Learning in Python},
  author  = {Pedregosa, Fabian and Varoquaux, Ga{\"e}l and Gramfort, Alexandre and Michel, Vincent and Thirion, Bertrand and Grisel, Olivier and Blondel, Mathieu and Prettenhofer, Peter and Weiss, Ron and Dubourg, Vincent and Vanderplas, Jake and Passos, Alexandre and Cournapeau, David and Brucher, Matthieu and Perrot, Matthieu and Duchesnay, {\'E}douard},
  journal = {J. Mach. Learn. Res.},
  volume  = {12},
  pages   = {2825--2830},
  year    = {2011}
}

@article{HongMandel1985,
  title = {Theory of parametric frequency down conversion of light},
  author = {Hong, C. K. and Mandel, L.},
  journal = {Phys. Rev. A},
  volume = {31},
  issue = {4},
  pages = {2409--2418},
  numpages = {0},
  year = {1985},
  month = {Apr},
  publisher = {American Physical Society},
  doi = {10.1103/PhysRevA.31.2409},
  url = {https://link.aps.org/doi/10.1103/PhysRevA.31.2409}
}

@article{Rubi1994,
  title = {Two-photon entanglement in type-II parametric down-conversion},
  author = {Shih, Y. H. and Sergienko, A. V. and Rubin, Morton H. and Kiess, T. E. and Alley, C. O.},
  journal = {Phys. Rev. A},
  volume = {50},
  issue = {1},
  pages = {23--28},
  numpages = {0},
  year = {1994},
  month = {Jul},
  publisher = {American Physical Society},
  doi = {10.1103/PhysRevA.50.23},
  url = {https://link.aps.org/doi/10.1103/PhysRevA.50.23}
}

@article{Werner1989,
  author  = {Werner, Reinhard F.},
  title   = {Quantum states with Einstein-Podolsky-Rosen correlations admitting a hidden-variable model},
  journal = {Phys. Rev. A},
  volume  = {40},
  number  = {8},
  pages   = {4277--4281},
  year    = {1989},
  doi     = {10.1103/PhysRevA.40.4277}
}

@article{czerwinski2022selected,
  title={Selected concepts of quantum state tomography},
  author={Czerwinski, Artur},
  journal={Optics},
  volume={3},
  number={3},
  pages={268--286},
  year={2022},
  publisher={MDPI},
  doi={10.3390/opt3030026}
}

@incollection{hasinoff2021photon,
  title={Photon, poisson noise},
  author={Hasinoff, Samuel W},
  booktitle={Computer vision: a reference guide},
  pages={980--982},
  year={2021},
  publisher={Springer}
}

@book{paris2004quantum,
  title={Quantum state estimation},
  author={Paris, Matteo and Rehacek, Jaroslav},
  volume={649},
  year={2004},
  publisher={Springer Science \& Business Media}
}

@article{hradil1997quantum,
  title={Quantum-state estimation},
  author={Hradil, Zdenek},
  journal={Phys. Rev. A},
  volume={55},
  number={3},
  pages={R1561},
  year={1997},
  publisher={APS}
}

@article{jamiolkowski1983minimal,
  title={Minimal number of operators for observability of N-level quantum systems},
  author={Jamio{\l}kowski, Andrzej},
  journal={Int. J. Theor. Phys.},
  volume={22},
  number={4},
  pages={369--376},
  year={1983},
  publisher={Springer}
}

@article{jamiolkowski2000complete,
  title={On complete and incomplete sets of observables, the principle of maximum entropy—revisited},
  author={Jamio{\l}kowski, Andrzej},
  journal={Rep. Math. Phys.},
  volume={46},
  number={3},
  pages={469--482},
  year={2000},
  publisher={Elsevier}
}

\appendix

\section{Generation of Datasets for Classification and Clustering with Bell States}
\label{appendixA}

\subsection{Generation of Maximally Entangled States}

The entangled-state ensemble was generated from the Bell states

\begin{equation}
|\Psi^{-}\rangle=\frac{1}{\sqrt{2}}
\left(
|01\rangle-|10\rangle
\right),
\end{equation}

and

\begin{equation}
|\Phi^{+}\rangle=\frac{1}{\sqrt{2}}
\left(
|00\rangle+|11\rangle
\right).
\end{equation}

To obtain a geometrically diverse collection of maximally entangled states, local unitary transformations were applied according to

\begin{equation}
U(\alpha,\beta,\gamma)=e^{-i\alpha\sigma_z/2}
e^{-i\beta\sigma_y/2}
e^{-i\gamma\sigma_z/2},
\end{equation}
where $\sigma_x$, $\sigma_y$, and $\sigma_z$ denote the Pauli matrices.

The transformation acting on the composite system was chosen as

\begin{equation}
U_{\mathrm{loc}}=\mathbb{I}_2
\otimes
U(\alpha,\beta,\gamma).
\end{equation}

Since local unitary operations preserve entanglement, varying the Euler angles $(\alpha,\beta,\gamma)$ generates physically distinct quantum states while maintaining maximal entanglement. From the resulting ensemble, a random subset containing $10\,000$ states was selected for further analysis.

\subsection{Generation of Separable States}

The separable-state ensemble was constructed from tensor products of single-qubit density matrices. Each qubit was parametrized using the Bloch-sphere representation

\begin{align}
r_x &= r \sin\theta \cos\phi, \\
r_y &= r \sin\theta \sin\phi, \\
r_z &= r \cos\theta.
\end{align}

The corresponding density matrix is given by

\begin{equation}
\rho(r,\theta,\phi)=\frac{1}{2}
\left(
\mathbb{I}_2
+
r_x\sigma_x
+
r_y\sigma_y
+
r_z\sigma_z
\right).
\end{equation}

Two-qubit separable states were generated according to

\begin{equation}
\rho_{\mathrm{sep}}=\rho_1\otimes\rho_2.
\end{equation}

To include states with different degrees of purity, the Bloch-vector length was selected from the set

\begin{equation}
r \in \{0.4,\,0.8,\,1.0\}.
\end{equation}

The angular parameters $\theta$ and $\phi$ were sampled over a numerical grid covering the Bloch sphere. This procedure produced more than $10^5$ valid separable states. A random subset of $10\,000$ states was subsequently selected for the ML analysis.

\subsection{Generation of Noisy Measurement Data}

For each density matrix $\rho$, the probability associated with the $k$-th SIC-POVM outcome was calculated according to Born's rule,

\begin{equation}
p_k=\mathrm{Tr}(M_k \rho).
\end{equation}

The measurement outcomes were generated with Poissonian shot noise,
\begin{equation}
n_k\sim \mathrm{Poisson}(N p_k),
\end{equation}
where
\begin{equation}
N  = 40,000
\end{equation}
denotes the average number of detected photon pairs per measurement setting.

Consequently, each quantum state was represented by a sixteen-dimensional measurement vector

\begin{equation}
\mathbf{x}=
(n_1,n_2,\ldots,n_{16}),
\end{equation}
whose components contain statistical fluctuations arising from photon shot noise.

The final dataset consisted of $10,000$ separable states and $10,000$ maximally entangled states. For supervised classification, each measurement vector was assigned a binary label

\begin{equation}
y =
\begin{cases}
0, & \text{separable state},\\
1, & \text{entangled state}.
\end{cases}
\end{equation}

The resulting dataset was subsequently employed for both classification and clustering analyses presented in the main text in Sec.~\ref{sec:Separable and Maximally Entangled States}.

\section{Generation of the Werner-State Dataset for Classification and Regression}
\label{appendixB}

\subsection{Werner States as a Source of Variable Entanglement}

The regression task considered in this work requires quantum states exhibiting continuously varying degrees of entanglement. For this purpose, we employed the family of Werner states, which provides a convenient interpolation between maximally mixed and maximally entangled two-qubit states.

The construction starts from the singlet Bell state
\begin{equation}
|\Psi^{-}\rangle=\frac{1}{\sqrt{2}}
\left(
|01\rangle-|10\rangle
\right).
\end{equation}

The corresponding density matrix is
\begin{equation}
\rho_{\Psi^-}=|\Psi^{-}\rangle
\langle\Psi^{-}|.
\end{equation}

The Werner state is then defined as
\begin{equation}\label{Wernerstate}
\rho_W(p)= p\,\rho_{\Psi^-} + (1-p)\frac{\mathbb{I}_4}{4},
\qquad
0 \leq p \leq 1,
\end{equation}
where $\mathbb{I}_4$ denotes the identity operator acting on the two-qubit Hilbert space.

The parameter $p$ controls the degree of mixing. For $p=0$, the state reduces to the maximally mixed state,
\begin{equation}
\rho_W(0)=\frac{\mathbb{I}_4}{4}.
\end{equation}

For $p=1$, the state becomes the maximally entangled singlet state,
\begin{equation}
\rho_W(1)=\rho_{\Psi^-}.
\end{equation}

Thus, the Werner family provides a simple one-parameter class of states ranging from a fully mixed state to a maximally entangled state.

\subsection{Local Unitary Transformations}

The Werner parameter $p$ controls the amount of entanglement. However, using only the canonical Werner form would restrict the generated states to a highly symmetric subset of the full two-qubit state space. To increase the geometrical diversity of the dataset, local unitary transformations were applied independently to both qubits.

The single-qubit unitary operator was parametrized as
\begin{equation}
U(\alpha,\beta,\gamma)=\begin{pmatrix}
e^{i\alpha}\cos\beta
&
e^{i\gamma}\sin\beta \\-e^{-i\gamma}\sin\beta
&
e^{-i\alpha}\cos\beta
\end{pmatrix}.
\label{eq:su2_parametrization}
\end{equation}

Two independent local unitary operators were defined as
\begin{equation}
U_A=U(\alpha_1,\beta_1,\gamma_1),
\end{equation}
and
\begin{equation}
U_B= U(\alpha_2,\beta_2,\gamma_2).
\end{equation}

The locally transformed Werner state was then generated according to
\begin{equation}
\rho_W^{(\mathrm{LU})}=(U_A\otimes U_B)
\rho_W(p)
(U_A\otimes U_B)^\dagger.
\label{eq:local_unitary_werner}
\end{equation}

Local unitary operations preserve entanglement measures. Consequently, states generated from the same value of $p$ have the same concurrence but may lead to different measurement statistics. This procedure therefore increases the diversity of measurement vectors while preserving controlled values of the target entanglement measure.

The parameters
\begin{equation}
p,\alpha_1,\beta_1,\gamma_1,\alpha_2,\beta_2,\gamma_2
\end{equation}
were sampled on a multidimensional numerical grid. This procedure produced
\begin{equation}
4,408,236
\end{equation}
valid two-qubit states. From this ensemble, a random subset of
\begin{equation}
250,000
\end{equation}
states was selected for further analysis of ML methods in Werner states.

\subsection{Concurrence for Two-Qubit States}\label{Concurrence}

For each generated quantum state, the amount of entanglement was quantified using concurrence. For an arbitrary two-qubit density matrix $\rho$, the spin-flipped state is defined as
\begin{equation}
\widetilde{\rho}=(\sigma_y\otimes\sigma_y)
\rho^{*}
(\sigma_y\otimes\sigma_y),
\label{eq:spin_flipped_state}
\end{equation}
where $\rho^{*}$ denotes complex conjugation in the computational basis, and
\begin{equation}
\sigma_y=\begin{pmatrix}
0 & -i \\
i & 0
\end{pmatrix}
\end{equation}
is the Pauli-$Y$ matrix.

Next, the positive operator
\begin{equation}
R=\sqrt{
\sqrt{\rho}
\widetilde{\rho}
\sqrt{\rho}
}
\label{eq:R_operator_concurrence}
\end{equation}
is constructed.

Let the eigenvalues of $R$ be ordered as
\begin{equation}
\lambda_1
\geq
\lambda_2
\geq
\lambda_3
\geq
\lambda_4.
\end{equation}

The concurrence is then given by
\begin{equation}
C(\rho)=\max
\left(0,\,\lambda_1-\lambda_2-\lambda_3-\lambda_4\right).
\label{eq:general_concurrence}
\end{equation}

Concurrence satisfies
\begin{equation}
0\leq C(\rho) \leq1,
\end{equation}
where
\begin{equation}
C(\rho)=0
\end{equation}
corresponds to separable states, while
\begin{equation}
C(\rho)=1
\end{equation}
corresponds to maximally entangled states.

For Werner states, the expression for concurrence can be simplified analytically. For the state defined in Eq.~\eqref{Wernerstate}, one obtains
\begin{equation}
C\left(\rho_W(p)\right)=\max
\left( 0,\frac{3p-1}{2}
\right).
\label{eq:werner_concurrence}
\end{equation}

This expression shows that Werner states are separable for
\begin{equation}
0
\leq
p
\leq
\frac{1}{3},
\end{equation}
and entangled for
\begin{equation}
\frac{1}{3}
<
p
\leq
1.
\end{equation}

Moreover, since local unitary transformations do not change concurrence, the locally transformed states in Eq.~\eqref{eq:local_unitary_werner} satisfy
\begin{equation}
C\left(\rho_W^{(\mathrm{LU})}\right)=C\!\left(\rho_W(p)\right).
\end{equation}

In the numerical implementation, concurrence was evaluated using the general formula in Eq.~\eqref{eq:general_concurrence}. The closed-form expression in Eq.~\eqref{eq:werner_concurrence} provides an analytical consistency check and confirms that the dataset covers the full range of concurrence values from $0$ to $1$.

\subsection{Generation of Noisy SIC-POVM Measurement Data}

For each locally transformed Werner state, the SIC-POVM measurement operators introduced in Section~\ref{sec:SIC-POVMData} were used to generate pseudo-experimental measurement data.

For the $k$-th measurement operator $M_k$, the corresponding probability was evaluated using Born's rule:
\begin{equation}
p_k = \mathrm{Tr}
\left(M_k \rho_W^{(\mathrm{LU})}
\right),
\qquad k=1,\ldots,16.
\label{eq:sic_probability_regression}
\end{equation}

To emulate finite photon-counting statistics, the measured count associated with the $k$-th outcome was sampled from a Poisson distribution:
\begin{equation}
n_k \sim \mathrm{Poisson}
\left(N p_k \right),
\label{eq:poisson_counts_regression}
\end{equation}
where
\begin{equation}
N=40,000
\end{equation}
denotes the average number of detected photon pairs per measurement setting.

This sampling procedure introduces shot-noise fluctuations and converts ideal probabilities into pseudo-experimental coincidence-count data.

\subsection{Construction of the Regression Dataset}

For each generated quantum state, the final data record consisted of sixteen noisy SIC-POVM counts and the corresponding concurrence value:
\begin{equation}
\rho_W^{(\mathrm{LU})} \hspace{0.35cm} \longleftrightarrow \hspace{0.35cm} \left(n_1, n_2, \ldots, n_{16},C,\mathcal{C}^{\mathrm{bin}},p\right).
\end{equation}

Equivalently, the ML input vector was
\begin{equation}
\mathbf{x}=\left(n_1,
n_2, \ldots,n_{16}\right),
\end{equation}
where the regression target was $y=C$ and for classification the output variable was $y = \mathcal{C}^{\mathrm{bin}}$.

The purpose of the regression task was therefore to learn the nonlinear mapping
\begin{equation}
f_{\mathrm{ML}}: \mathbb{R}^{16} \rightarrow [0,1],
\end{equation}
such that
\begin{equation}
f_{\mathrm{ML}}(\mathbf{x})
\approx
C(\rho_W^{(\mathrm{LU})}).
\end{equation}

This dataset was used to evaluate whether ML algorithms can estimate the amount of quantum entanglement directly from noisy measurement statistics, without reconstructing the underlying density matrix.

The resulting dataset was subsequently employed for both classification and regression analyses of Werner states presented in the main text in Sec.~\ref{sec:Werner States}.

\end{document}